\documentclass[
    ,final            
  ]
  {aipproc}

\usepackage[mathscr]{eucal}
 \usepackage{amsmath, amssymb}

  \def\sun{\hbox{$\odot$}}
   
\newcommand{\be}{\begin{equation}}
\newcommand{\ee}{\end{equation}}
\newcommand{\bdm}{\begin{displaymath}}
\newcommand{\edm}{\end{displaymath}}
\newcommand{\vect}[1]{\mathbf{#1}}
\def\la{\mathrel{\mathchoice {\vcenter{\offinterlineskip\halign{\hfil
$\displaystyle##$\hfil\cr<\cr\sim\cr}}}
{\vcenter{\offinterlineskip\halign{\hfil$\textstyle##$\hfil\cr
<\cr\sim\cr}}}
{\vcenter{\offinterlineskip\halign{\hfil$\scriptstyle##$\hfil\cr
<\cr\sim\cr}}}
{\vcenter{\offinterlineskip\halign{\hfil$\scriptscriptstyle##$\hfil\cr
<\cr\sim\cr}}}}}

\layoutstyle{6x9}

\begin{document}

\noindent {\it to appear in Astronomy Reports, July 2012}

\vspace{1cm}

\title{Signs of magnetic accretion in X-ray pulsars}


\author{N.R.\,Ikhsanov}{
  address={Pulkovo Observatory, Pulkovskoe Shosse 65, Saint-Petersburg 196140, Russia}
}

\author{N.G.\,Beskrovnaya}{
  address={Pulkovo Observatory, Pulkovskoe Shosse 65, Saint-Petersburg 196140, Russia}
}

\begin{abstract}
The spin-down mechanism of accreting neutron stars is discussed with an application to one of the best studied X-ray pulsars GX\,301--2. We show that the maximum possible spin-down torque applied to a neutron star from the accretion flow can be evaluated as $K_{\rm sd}^{\rm (t)} = \mu^2/\left(r_{\rm m} r_{\rm cor}\right)^{3/2}$. The spin-down rate of the neutron star in GX\,301--2 can be explained provided the magnetospheric radius of the neutron star is smaller than its canonical value. We calculate the magnetospheric radius considering the mass-transfer in the binary system in the frame of the magnetic accretion scenario suggested by V.F.\,Shvartsman. The spin-down rate of the neutron star expected within this approach is in a good agreement with that derived from observations of GX\,301--2.
\end{abstract}

\maketitle


  \section{Introduction}

 Observations of the long-period ($P_{\rm s} \simeq 685$\,s) X-ray pulsar GX\,301--2 have shown periods (lasting up to a few years) when
the neutron star spin frequency ($\nu = 1/P_{\rm s}$) is steadily  decreasing at the rate  $\dot{\nu}_0 \simeq -10^{-13}\,{\rm Hz\,s^{-1}}$ \cite{Doroshenko-etal-2010}. This pulsar has been identified with a neutron star in a High Mass X-ray Binary with the orbital period  $P_{\rm orb} \simeq 41.5$\,d. The massive component is the early-type supergiant (Wray\,977 \cite{Sato-etal-1986}). It underfills its Roche lobe and looses material at the rate  $\dot{M}_{\rm out} \simeq 10^{-5}\,{\rm M_{\sun}\,yr^{-1}}$ in a form of relatively slow, $v_{\rm w} \sim 300-400\,{\rm km\,s^{-1}}$, stellar wind \cite{Kaper-etal-2006}. The neutron star moving in the wind of its companion captures material and accretes onto its surface at the rate:
 \be
\dot{M}_{\rm a} = \frac{L_{\rm X} R_{\rm ns}}{GM_{\rm ns}} \sim 5 \times 10^{16}\ L_{37}\,R_6 m^{-1}\,{\rm g\,s^{-1}},
 \ee
 where $R_6 = R_{\rm ns}/10^6$\,cm and $m = M_{\rm ns}/1.4\,M_{\sun}$ are the radius and mass of the neutron star, and  $L_{37} = L_{\rm X}/10^{37}\,{\rm erg\,s^{-1}}$ is the X-ray luminosity of the pulsar normalized according to  \cite{Chichkov-etal-1995, Kaper-etal-2006}.  Observations have revealed no signs of developed Keplerian accretion disk in the system. Besides, the relative velocity of the neutron star in the wind of its massive companion,  $\vect{v}_{\rm rel} = \vect{v}_{\rm w} + \vect{v}_{\rm ns}$, exceeds the upper limit  (see \cite{Ikhsanov-2007}),
 \be\label{vcr}
v_{\rm cr} \simeq 200\ \xi_{0.2}^{1/4}\ \mu_{30}^{-1/14}\ m^{11/28}\ \dot{M}_{17}^{1/28}\ \left(\frac{P_{\rm orb}}{41.5\,{\rm d}}\right)^{-1/4}\  {\rm km\,s^{-1}},
 \ee
at which the angular momentum of the captured material is sufficient to form a Keplerian disk. Therefore, the accretion process in
GX\,301--2 is usually treated in quasi-spherical approximation. Here $v_{\rm ns} \sim 250\,{\rm km\,s^{-1}}$ is the linear orbital velocity of the neutron star, $\mu_{30}$ is its dipole magnetic moment in units of $10^{30}\,{\rm G\,cm^3}$, $\dot{M}_{17}$ is the rate at which the neutron star captures material from the stellar wind in units of $10^{17}\,{\rm g\,s^{-1}}$ and $\xi_{0.2} = \xi/0.2$ is a parameter accounting for the angular momentum dissipation   due to inhomogeneities in the accretion flow. It is normalized to its average value according to results of numerical simulations of wind-fed accretion in the approximation of non-magnetized accretion flow  (see \cite{Ruffert-1999} and references therein).

Interpretation of the observed spin-down of the neutron star in the frame of popular scenarios of spherical accretion encounters some difficulties. Lipunov  (see \cite{Lipunov-1982, Lipunov-1992}) has shown that the spin-down torque applied to the neutron star undergoing quasi-spherical accretion can be evaluated from the expression  $K_{\rm sd}^{\rm (s)} = k_{\rm t} \mu^2/r_{\rm cor}^3$. Here  $r_{\rm cor} = \left(GM_{\rm ns}/\omega_{\rm s}^2\right)^{1/3}$ is the corotation radius of the neutron star spinning at angular velocity  $\omega_{\rm s} = 2 \pi/P_{\rm s}$, and $k_{\rm t} <1$ is the efficiency parameter. The spin-down rate of the neutron star in  GX\,301--2 in this case is limited as  $2 \pi I \dot{\nu}_{\rm s} \la K_{\rm sd}^{\rm (s)}$ and its absolute value,
 \be\label{nus}
 \dot{\nu}_{\rm s} \leq 3 \times 10^{-16}\ k_{\rm t} I_{45}^{-1} m^{-1} R_6^6 P_{685}^{-2} \left(\frac{B_*}{B_{\rm CRSF}}\right)^2\ {\rm Hz\,s^{-1}},
 \ee
turns out to be a factor of 340 less than that derived from observations. Here $I_{45} = I/10^{45}\,{\rm g\,cm^2}$ is the moment of inertia of the neutron star, $P_{685}$ is its spin period in the units 685\,s and $B_{\rm CRSF} = 4 \times 10^{12}$\,G is the magnetic field strength on the  stellar surface estimated through observations of the cyclotron line in the X-ray spectrum of the pulsar \cite{La-Barbera-etal-2005}.

The value of spin-down torque in the scenarios accounting for an influence of the X-ray luminosity of the pulsar on the accretion flow \cite{Illarionov-Kompaneets-1990} and  a possibility for a hot gaseous envelope to form around the neutron star magnetosphere  \cite{Shakura-etal-2012}, under certain conditions can be as large as  $K_{\rm sd}^{\rm (t)} = z_0 \dot{M}_{\rm c} \omega_{\rm s} r_{\rm m}^2$. Here $\dot{M}_{\rm c} = \pi R_{\rm G}^2 \rho_{\infty} V_{\rm rel}$ is the mass of gas with which the neutron star interacts in a unit time moving through the wind of its companion, $R_{\rm G} = 2GM_{\rm ns}/V_{\rm rel}^2$ is the gravitational capture (Bondi) radius and  $\rho_{\infty}$ is the density of stellar wind at the Bondi radius. $r_{\rm m}$ is the radius of the neutron star magnetosphere which in the case of spherical accretion is  $r_{\rm m} \geq r_{\rm a}$, where \cite{Arons-Lea-1976}
 \be
 r_{\rm a} = \left(\frac{\mu^2}{\dot{M}_{\rm c} (2 GM_{\rm ns})^{1/2}}\right)^{2/7}
 \ee
 and $z_0 < 1$  is a numerical parameter. The expected spin-down rate of the neutron star in  GX\,301--2 in the frame of these scenarios
\be\label{nucr}
\dot{\nu}_{\rm t} \simeq 2 \times 10^{-14}\,{\rm Hz\,s^{-1}}\ z_0\,I_{45}^{-1}\,P_{685}^{-1}\,m^{-5/7}\,R_6^{27/7}\,L_{37}^{3/7}\,\left(\frac{B_*}{B_{\rm CRSF}}\right)^{8/7},
 \ee
exceeds an estimate derived from the canonical model of quasi-spherical accretion  (see expression ~\ref{nus}), but still remains a factor of 5 less than the observed value.

Analyzing this discrepancy, Doroshenko et al. \cite{Doroshenko-etal-2010} have assumed that the magnetic field strength on the surface of the neutron star in   GX\,301--2 is a factor of  30 higher than that estimated through observations of the cyclotron line in the X-ray spectrum of the source,  $B_{\rm CRSF}$. The neutron star under this assumption proves to be an accreting magnetar with the surface field in excess of $10^{14}$\,G. In this paper we argue that so dramatic revision of our notions of magnetic fields in accreting neutron stars is premature. An apparent contradiction between the predicted and observed spin-down rate of the neutron star indicates that the accretion picture in the long-period X-ray pulsars may differ from commonly adopted scenarios. In particular, the observed spin-down rate of the neutron star can be explained in the case of magnetic accretion   \cite{Shvartsman-1971} without additional assumptions about the value of the neutron star magnetic field.

     \section{Magnetic accretion}

As was first shown by Shvartsman \cite{Shvartsman-1971}, a picture of the wind-fed accretion in a massive binary strongly depends on the value of magnetic field in the matter captured by the compact star. The magnetic field in the free-falling accretion flow is dominated by the radial component. This is connected with the fact that in the process of spherical accretion the transverse scales contract as $\sim r^{-2}$, while the radial scales expand as $\sim r^{1/2}$ \cite{Zeldovich-Shakura-1969}.  Therefore, under the condition of the magnetic flux conservation
the field strength in the accretion flow can be approximated as  $B_{\rm r} \propto r^{-2}$ \cite{Bisnovatyi-Kogan-Fridman-1970}. This implies that the magnetic energy density,   $E_{\rm m} = B_{\rm r}^2/8 \pi$, in the free-falling material increases as it approaches a compact star as
 \be
 E_{\rm m}(r) = E_{\rm m}(R_{\rm G}) \left(\frac{R_{\rm G}}{r}\right)^4,
 \ee
while the kinetic energy of the accretion flow is
  \be
 E_{\rm ram}(r) = E_{\rm ram}(R_{\rm G}) \left(\frac{R_{\rm G}}{r} \right)^{5/2}.
 \ee
 Here $E_{\rm m}(R_{\rm G}) = \beta^{-1} E_{\rm th}(R_{\rm G})$ is the magnetic energy density in the accretion flow at the Bondi radius normalized by the thermal energy of the flow, $E_{\rm th}(R_{\rm G}) = \rho_{\infty} c_{\rm s}^2(R_{\rm G})$, with the  parameter $\beta$, and $c_{\rm s}(R_{\rm G})$ is the sound speed in the captured material. $E_{\rm ram}(R_{\rm G}) = \rho_{\infty} v_{\rm rel}^2$ is the kinetic energy density in the flow at the Bondi radius.

Then solving the equation $E_{\rm m}(R_{\rm sh}) = E_{\rm ram}(R_{\rm sh})$  we can find the distance $R_{\rm sh}$ (hereafter, Shvartsman radius) at which the magnetic energy density in the accretion flow reaches its kinetic energy density in the form
  \be\label{rsh}
 R_{\rm sh} = \beta^{-2/3} \left(\frac{c_{\rm s}}{v_{\rm rel}}\right)^{4/3} R_{\rm G} =
 \beta^{-2/3}\ \frac{2 GM_{\rm ns} c_{\rm s}^{4/3}}{v_{\rm rel}^{10/3}}.
 \ee
Shvartsman \cite{Shvartsman-1971} has pointed out that for the accretion flow in the spatial region  $r < R_{\rm sh}$ to proceed further the dissipation of its magnetic field should occur.  Otherwise, the magnetic energy in the flow would exceed its gravitational energy which contradicts the energy conservation law. Thus, the radial velocity of the accreting material in this region can be expressed as $v_{\rm r} \sim r/t_{\rm rec}$, where
 \be\label{trec}
 t_{\rm rec} = \frac{r}{\eta_{\rm m} V_{\rm A}} = \eta_{\rm m}^{-1}\ t_{\rm ff}\
 \left(\frac{V_{\rm ff}}{V_{\rm A}}\right)
 \ee
is the characteristic time of field dissipation due to magnetic reconnection. Here
 $V_{\rm A} = B_{\rm r}/(4 \pi \rho)^{1/2}$ is the Alfven velocity in the accretion flow,  $t_{\rm ff} = r/V_{\rm ff}$ and $V_{\rm ff}(r) = \left(GM_{\rm ns}/r\right)^{1/2}$ is the free-fall time and velocity. $\eta_{\rm m}$ is the efficiency parameter of the reconnection process. Its value  depends on the physical parameters of plasma and magnetic field configuration in the region of reconnection and  ranges  in the interval $0 < \eta_{\rm m} <0.1$ \cite{Kadomtsev-1987}. Since $V_{\rm A} \leq V_{\rm ff}$ (the equality  is met at the Shvartsman radius), the timescale  of field dissipation in the flow remains significantly less than the dynamical  (free-fall) time,
 $t_{\rm ff}$, during the entire accretion process.
This confirms the validity of assumption about the magnetic flux conservation in the free-falling material, but, on the other hand, this implies that  the flow is decelerated by its own magnetic field at the Shvartsman radius and accretion switches  into the diffusion regime.

Basic conclusions of the scenario suggested by Shvartsman have been later confirmed with quantitative assessments by Bisnovatyi-Kogan and Ruzmaikin \cite{Bisnovatyi-Kogan-Ruzmaikin-1974, Bisnovatyi-Kogan-Ruzmaikin-1976}, and by results  of numerical simulations of the spherical magnetic accretion presented in \cite{Igumenshchev-etal-2003, Igumenshchev-2006}. These authors have shown that magnetic field amplification in the free-falling material leads to deceleration of the accretion flow at  $\sim R_{\rm sh}$ and its shock-heating up to adiabatic temperature. Accretion inside the Shvartsman radius occurs on the timescale of magnetic field dissipation. However, the accretion picture essentially depends on the efficiency of cooling processes in the accreting material. If the cooling time at the
 Shvartsman radius, $t_{\rm cool}(R_{\rm sh})$, exceeds the heating time due to magnetic energy dissipation, $t_{\rm rec}(R_{\rm sh})$, accretion inside the  Shvartsman radius occurs from a hot turbulent envelope with some portion of matter leaving the system in a form of jets
\cite{Igumenshchev-etal-2003, Igumenshchev-2006}. Otherwise, the flow is transformed into the magnetic slab with parameters depending, in particular, on the geometry of the large-scale magnetic field in the accretion flow \cite{Bisnovatyi-Kogan-Ruzmaikin-1974, Bisnovatyi-Kogan-Ruzmaikin-1976}.

 \section{Magnetic accretion in X-ray pulsars}

A possibility for magnetic accretion to be realized in an X-ray pulsar is determined by relation between the Shvartsman radius and the radius of the neutron star magnetosphere. If  $R_{\rm sh} < r_{\rm a}$, the accretion flow reaches the magnetospheric boundary of the neutron star in the free-fall regime and the accretion process in this case can be considered in the quasi-spherical approximation. Otherwise, the accretion picture  should be treated in terms of magnetic accretion.
Solving inequality  $R_{\rm sh} > r_{\rm a}$ for $v_{\rm rel}$, we find that magnetic accretion in the massive X-ray binaries can occur provided  $v_{\rm rel} < v_{\rm mca}$, where
 \be
 v_{\rm mca} = \beta^{-1/5}\,(2GM_{\rm ns})^{12/35}\,\mu^{-6/35}\,\dot{M}_{\rm c}^{3/35}\ c_{\rm s}^{2/5}.
 \ee

For typical parameters of the long-period pulsars (see \cite{Ikhsanov-2007}), this yields
  \be
v_{\rm mca} \simeq 680\ \beta^{-1/5}\ m^{12/35} \mu_{30}^{-6/35}\ \dot{M}_{17}^{3/35}\ \left(\frac{c_{\rm s}}{10\,{\rm km\,s^{-1}}}\right)^{2/5}\ {\rm km\,s^{-1}}.
 \ee
Thus, for moderate values of the parameter  $\beta$,  $v_{\rm mca}$ significantly exceeds  $v_{\rm cr}$ (see Eq.~\ref{vcr}). This allows us to distinguish a subclass of X-ray pulsars in which the magnetic accretion scenario can be realized. This subclass is determined by the following condition $v_{\rm cr} < v_{\rm rel} < v_{\rm mca}$.

The parameter  $\beta$ in the material captured by the neutron star at the Bondi radius can be, in the first approximation, estimated as follows. Let $B_{\rm ms}$ denote the magnetic field strength on the surface of the massive counterpart to the neutron star. The field strength in the stellar wind ejected by this star is decreasing as it is moving away from the star as $B_{\rm w}(a) \sim B_{\rm ms} \left(R_{\rm ms}/a\right)^3$, up to the distance $a_{\rm k}$, at which the kinetic energy density of the wind becomes equal to the magnetic energy density of the stellar dipole field. Here  $R_{\rm ms}$ is the radius of the massive star, and $a$ is the distance from its center. The magnetic field strength in the stellar wind in the region  $a > a_{\rm k}$ decreases  $\propto a^{-2}$ \cite{Walder-etal-2011}. Thus, the magnetic energy density in the stellar wind in the vicinity of the neutron star  can be evaluated using the expression  $E_{\rm m}(a_0) = \mu_{\rm ms}^2/\left(2 \pi a_{\rm k}^2 a_0^4\right)$, where $a_0$ is the orbital separation. Putting the numerical values yields
  \be\label{bwa}
E_{\rm m}(a_0) \simeq 0.33\ {\rm erg\,cm^{-3}}\ a_{13}^{-4}  \left(\frac{\mu_{\rm ms}}{10^{39}\,{\rm G\,cm^3}}\right)^2 \left(\frac{a_{\rm k}}{100\,R_{\sun}}\right)^{-2},
 \ee
 where $a_{13} = a_0/10^{13}$\,cm, and $\mu_{\rm ms}$ is the dipole magnetic moment of the massive companion which was normalized according to spectropolarimetric data on massive stars \cite{Hubrig-etal-2006, Oksala-etal-2010, Martins-etal-2010}. These observations have shown that magnetic field strength of a few thousand Gauss is not unique among early-type stars.

The thermal energy density in the material captured by the neutron star at the Bondi radius,  $E_{\rm th}(R_{\rm G}) = \rho_{\infty} c_{\rm s}^2$, can be expressed by taking into account that $\rho_{\infty} = \dot{M}_{\rm c}/\pi R_{\rm G}^2 v_{\rm rel}$ as follows,
\be
E_{\rm th} \simeq 0.02\,{\rm erg\,cm^{-3}}\ m^{-2}\ \dot{M}_{17}\ \left(\frac{v_{\rm rel}}{500\,{km\,s^{-1}}}\right)^3 \left(\frac{v_{\rm s}}{10\,{\rm km\,s^{-1}}}\right)^2
 \ee
These estimates demonstrate that the parameter $\beta$ in the stellar wind captured by the neutron star can be of the order of or even less that unity in not too wide pairs with a moderate stellar wind velocity. It should be noted that $\beta \sim 1$ has been measured in the solar wind plasma in the vicinity of  the Earth. This indirectly confirms the validity of assumption about significant magnetization of stellar wind adopted within magnetic accretion scenario.

Finally, transformation of the spherical flow to the magnetic slab inside the Shvartsman radius
\cite{Bisnovatyi-Kogan-Ruzmaikin-1974, Bisnovatyi-Kogan-Ruzmaikin-1976} can occur provided $t_{\rm cool} < t_{\rm rec}$. The most effective cooling mechanism in the accretion flow in X-ray pulsars is the inverse Compton scattering X-ray photons emitted from the neutron star surface on the hot electrons of plasma surrounding its magnetosphere \cite{Arons-Lea-1976, Elsner-Lamb-1977}. Solving inequality  $t_{\rm c}(R_{\rm sh}) < t_{\rm rec}$, where
 \be\label{tcomp}
t_{\rm c}(r) = \frac{3 \pi\,r^2\,m_{\rm e}\,c^2}{2\,\sigma_{\rm T}\,L_{\rm X}}
 \ee
is the Compton cooling time \cite{Elsner-Lamb-1977}, we come to the conclusion that a formation of the magnetic slab surrounding the neutron star magnetosphere is possible  only if the X-ray luminosity of the pulsar exceeds the critical value: $L_{\rm X} > L_{\rm cr}$, where
\be
 L_{\rm cr} \simeq 3 \times 10^{33}\ \mu_{30}^{1/4}\ m^{1/2}\ R_6^{-1/8}\ \left(\frac{\eta_{\rm m}}{0.001}\right) \left(\frac{R_{\rm sh}}{r_{\rm a}}\right)^{1/2}\ {\rm erg\,s^{-1}}.
 \ee
Here $m_{\rm e}$ is the electron mass, and $\sigma_{\rm T}$ is the Thomson cross-section.

The plasma density at the inner radius of the slab, $\rho_{\rm sl}$, can be evaluated taking into account that the thermal pressure in the slab is equal to the magnetic pressure at the magnetospheric radius as
 \be\label{rhosl}
  \rho_{\rm sl} = \frac{\mu^2\,m_{\rm p}}{2 \pi\,k_{\rm B}\,T_0\,r_{\rm m}^6}.
  \ee
Here $m_{\rm p}$ and $k_{\rm B}$ are the proton mass and Boltzmann constant, and $T_0$ is the gas temperature at the inner radius of the magnetic slab.

 \section{Evolution of the neutron star spin period}

The accretion picture in  GX\,301--2 differs from that expected within the quasi-spherical accretion scenario by exceptionally high rate of angular momentum dissipation in the accretion flow. We  came to this conclusion taking into account that spin-down of the neutron star undergoing quasi-spherical accretion is possible only provided the angular velocity of the accreting material at the magnetospheric boundary, $\omega_{\rm em}(r_{\rm m}) = \xi \Omega_{\rm orb} \left(R_{\rm G}/r_{\rm m}\right)^2$, is less than angular velocity of the star itself, $\omega_{\rm s}$ (see \cite{Bisnovatyi-Kogan-1991}). Here $\Omega_{\rm orb} = 2 \pi/P_{\rm orb}$ is the average value of angular orbital velocity.  Solving inequality  $\omega_{\rm em}(r_{\rm m}) < \omega_{\rm s}$ for $\xi$, we find
  \begin{eqnarray}\label{ksd}
\xi & < & 0.03~m^{-12/7}\ L_{37}^{-4/7}\ R_6^{20/7}\ \left(\frac{P_{\rm orb}}{41.5\,{\rm d}}\right) ~ \times\  \\
    \nonumber
 & & \times\  \left(\frac{P_{\rm s}}{685\,{\rm s}}\right)^{-1} \left(\frac{v_{\rm rel}}{400\,{\rm km\,s^{-1}}}\right)^4 \left(\frac{B_*}{B_{\rm CRSF}}\right)^{8/7}.
  \end{eqnarray}
This value of  $\xi$ is almost an order of magnitude smaller than the average value of this parameter evaluated in numerical calculations of wind-fed accretion made under assumption  $\beta \gg 1$ (see \cite{Ruffert-1999} and references therein).

Significant dissipation of the angular momentum in the accretion flow proves to be possible if the neutron star is accreting material from the turbulent quasi-statical envelope  \cite{Davies-Pringle-1981, Shakura-etal-2012} or if accreting material possesses strong enough magnetic field  \cite{Mestel-1959}. The spin-down torque applied to the neutron star from the accretion flow can be estimated as  $K_{\rm sd}^{\rm (t)} = k_{\rm t} \dot{M}_{\rm c} \omega_{\rm s} r_{\rm m}^2$, which can be written by substituting  $\dot{M}_{\rm c} = \mu^2/\left(2 GM_{\rm ns} r_{\rm m}^7 \right)^{1/2}$ in the form
 \be\label{ksdt}
K_{\rm sd}^{\rm (t)} = \frac{k_{\rm t}\,\mu^2\,\omega_{\rm s}}{r_{\rm m}^{3/2} (2 GM_{\rm ns})^{1/2}} =  \frac{k_{\rm t}\,\mu^2}{\left(r_{\rm m}\,r_{\rm cor}\right)^{3/2}}.
 \ee

As seen from this expression, the spin-down torque applied to the neutron star from the accretion flow essentially depends on its magnetospheric radius and is the greater the less the value of magnetospheric radius is.  If a neutron star accretes material from a quasi-spherical flow or a hot turbulent envelope, its magnetospheric radius is  $r_{\rm m} \sim r_{\rm a}$ \cite{Arons-Lea-1976, Davies-Pringle-1981, Shakura-etal-2012}. However, the spin-down rate of the neutron star in this case turns out to be insufficient to explain the observed breaking of  GX\,301--2 (see expression~\ref{nucr}).

The magnetospheric radius of the neutron star accreting material from the magnetic slab depends on the mechanism of plasma penetration into the stellar magnetic field. If it occurs due to interchange instabilities of the magnetospheric boundary (Rayleigh-Tailor and Kelvin-Helmholtz instabilities), then, similar to the previous case, the magnetospheric radius will be close to its canonical value,  $r_{\rm a}$. If the interchange instabilities of the magnetospheric boundary are suppressed, the plasma penetration into the magnetic field occurs due to diffusion process governed by magnetic reconnection at the rate  \cite{Elsner-Lamb-1984}:
 \be\label{dmfin-1}
 \dot{M}_{\rm in}(r_{\rm m}) = 4 \pi r_{\rm m} \delta_{\rm m} \rho_0 V_{\rm ff}(r_{\rm m}) = 4 \pi r_{\rm m}^{3/2} D_{\rm eff}^{1/2} \rho_0 V_{\rm ff}^{1/2}(r_{\rm m}).
  \ee
 Here $\delta_{\rm m} = \left(D_{\rm eff}\ \tau_{\rm d}\right)^{1/2}$ is the depth of the diffusion layer on the magnetospheric boundary (magnetopause), $D_{\rm eff}$ is the effective diffusion coefficient and $\rho_0$ is the plasma density at the boundary. The characteristic time of plasma penetration  into the stellar field,  $\tau_{\rm d}$, is determined by the time on which the plasma having diffused into the field leaves the magnetopause moving along the magnetic field lines due to gravitational attraction by the neutron star. Under the conditions of interest this time corresponds to the dynamical (free-fall) time at the magnetospheric boundary, i.e.  $\tau_{\rm d} \sim t_{\rm ff}(r_{\rm m})$ \cite{Elsner-Lamb-1984}.

 It is necessary to note that the situation in which the interchange instabilities of the magnetospheric boundary are suppressed due to magnetic shear generation in the magnetopause (see  \cite{Ikhsanov-Pustilnik-1996, Anzer-Boerner-1983, Malagoli-etal-1996}) is not exceptional. The high efficiency of this stabilization is well-known, in particular, from laboratory experiments with TOKAMAKs  \cite{Kadomtsev-Shafranov-1983}. Besides, studies of the Earth magnetosphere have shown that the rate at which the solar wind plasma penetrates into its magnetic field can be well approximated by the expression~(\ref{dmfin-1}), in which $D_{\rm eff}$ is the Bohm diffusion coefficient,
\be\label{dbohm}
D_{\rm B} = \alpha_{\rm B} \frac{c k_{\rm B} T_0}{2 e B(r_{\rm m})},
 \ee
and the efficiency parameter  $\alpha_{\rm B}$ ranges between  $0.1-0.25$ \cite{Gosling-etal-1991}. Here $e$ is the electron charge, $T_0$ is the plasma temperature, and  $B(r_{\rm m})$ is the magnetic field strength at the magnetospheric boundary. This provides us with additional grounds to assume that plasma penetration into the magnetosphere of the neutron star can occur due to Bohm diffusion.

The magnetospheric radius of the neutron star in this case can be evaluated by taking into account the stationary character of the accretion process, that is adopting the rate of plasma penetration into the stellar magnetic field equal to the  mass  capture rate  by the neutron star from the wind of its companion as well as to the rate of plasma accretion onto the neutron star surface , i.e. $\dot{M}_{\rm in}(r_{\rm mca}) = L_{\rm X} R_{\rm ns}/GM_{\rm ns}$.      Solving this equation, we find
 \be\label{rmb}
r_{\rm mca} \simeq 8 \times 10^7\,{\rm cm}\ \times \ \alpha_{0.1}^{2/13}\ \mu_{30}^{6/13}\ T_6^{-2/13}\ m^{5/13}\ L_{37}^{-4/13}\ R_6^{-4/13},
  \ee
 where $\alpha_{0.1}=\alpha_{\rm B}/0.1$ and $T_6 = T_0/10^6$\,K is the plasma temperature at the inner radius of the magnetic slab normalized according to the observational results presented in \cite{Masetti-etal-2006}. The spin-down torque applied to the neutron star from the magnetic slab turns out to be
 \be
K_{\rm sd}^{\rm (sl)} = \frac{k_{\rm t}\,\mu^2}{\left(r_{\rm mca}\,r_{\rm cor}\right)^{3/2}}.
 \ee
This result makes it possible to estimate an expected rate of the neutron star spin-down within the magnetic accretion scenario using the expression $\dot{\nu}_{\rm sd}^{\rm (mca)} = K_{\rm sd}^{\rm (sl)}/2 \pi I$. Substituting the parameters of the pulsar GX\,301--2, we get:
  \begin{eqnarray}\label{ksd1}
\dot{\nu}_{\rm sd}^{\rm (mca)} & \simeq & 7 \times 10^{-13}\,{\rm Hz\,s^{-1}} ~ k_{\rm t}\ \alpha_{0.1}^{-3/13}\ m^{-14/13}\ I_{45}^{-1}\ T_6^{3/13}\ L_{37}^{6/13} \\
     \nonumber
 & & \times \ R_6^{57/13} \left(\frac{P_{\rm s}}{685\,{\rm s}}\right)^{-1}  \left(\frac{B_*}{B_{\rm CRSF}}\right)^{17/13}.
  \end{eqnarray}
Thus, the observed spin-down rate of the neutron star in the X-ray pulsar GX\,301--2 can be explained in terms of the magnetic accretion scenario provided $k_{\rm t} \geq 0.14$.

 \section{Conclusions}

The magnetic accretion scenario has been up to now applied exclusively to the case of accretion on to the black holes (see, e.g.  \cite{Shvartsman-1971, Bisnovatyi-Kogan-Ruzmaikin-1974, Bisnovatyi-Kogan-Ruzmaikin-1976, Igumenshchev-etal-2003, Igumenshchev-2006}). It is difficult to perform observational checks  of main predictions of these studies. Our paper presents one of the first attempts to apply the magnetic accretion scenario to reconstruction of the mass exchange picture between the components of the X-ray binary system as well as for interpretation of observational appearance of these objects. The most important result of our study is the conclusion that the spin-down rate of the neutron star undergoing magnetic accretion proves to be higher than in case of stars accreting non-magnetized material. Observational verification of this result does not encounter any difficulties nowadays. This is illustrated by presented above explanation of exceptionally high spin-down rate of the neutron star in  GX\,301--2. This result, however, points out a necessity to analyze in more detail the mechanisms used in the magnetic accretion scenario which remained poorly understood so far. It is worthwhile to note  that the influence of the magnetic field of the accretion flow on both  its structure and parameters of the neutron star magnetosphere can lead to increase of the rate of matter outflow from the magnetospheric boundary considered in the paper by Lovelace et al. \cite{Lovelace-etal-1995}.The neutron star spin-down rate in this case can be higher. Besides, investigating the parameters of outflowing matter by means of X-ray spectroscopy opens additional opportunity to identify the magnetic accretion scenario.

\begin{theacknowledgments}
The authors are grateful to L.A.\,Pustil'nik for interesting discussions and useful comments. This work was supported by the Program of Presidium of Russian Academy of Sciences N\,21, and NSH-1625.2012.2.
\end{theacknowledgments}

\end{document}